 \documentclass[]{emulateapj}
\pdfoutput=1

\usepackage{amsbsy}

\shorttitle{Coherent structures in photospheric flows}
\shortauthors{Chian {\it et al.}}

\begin{document}

\title{Detection of coherent structures in photospheric turbulent flows}


\author{Abraham C.-L. Chian\altaffilmark{1,}\altaffilmark{2}}
\email{abraham.chian@gmail.com}

\author{Erico L. Rempel\altaffilmark{1,}\altaffilmark{3}}
\email{rempel@ita.br}

\author{Guillaume Aulanier\altaffilmark{2}}

\author{Brigitte Schmieder\altaffilmark{2}}

\author{Shawn C. Shadden\altaffilmark{4}}

\author{Brian T. Welsch\altaffilmark{5}}


\author{Anthony R. Yeates\altaffilmark{6}}


\altaffiltext{1}{National Institute for Space Research (INPE), World Institute for 
Space Environment Research (WISER), P.O. Box 515, 12227-010 S\~ao Jos\'e dos Campos -- SP, Brazil}
\altaffiltext{2}{Observatoire de Paris, LESIA, CNRS, 92190 Meudon, France}
\altaffiltext{3}{Institute of Aeronautical Technology (ITA), WISER, 12228--900 S\~ao Jos\'e dos Campos -- SP, Brazil}
\altaffiltext{4}{Department of Mechanical Engineering, University of California, Berkeley, CA 94720, USA}
\altaffiltext{5}{Space Science Laboratory, University of California, Berkeley, CA 94720, USA}
\altaffiltext{6}{Department of Mathematical Sciences, Durham University, Durham, DH1 3LE, UK}

\begin{abstract}
We study coherent structures in solar photospheric flows in a plage in the vicinity of the active region AR 10930 using the horizontal velocity data derived from Hinode/SOT magnetograms. Eulerian and Lagrangian coherent structures are detected by computing the Q-criterion and the finite-time Lyapunov exponents of the velocity field, respectively. Our analysis indicates that, on average, the deformation
Eulerian coherent structures dominate over the vortical Eulerian coherent structures in the plage region.
We demonstrate the correspondence of the network of high magnetic flux concentration to the attracting Lagrangian
coherent structures (a-LCS) in the photospheric velocity based on both observations and numerical simulations. 
In addition, the computation of a-LCS provides a measure of the local rate of contraction/expansion of the flow.
\end{abstract}



\maketitle

\section{Introduction}

The spatiotemporal patterns and transport of magnetic field in the photosphere are driven by the turbulent plasma flows in the solar convection zone. Solar magnetic fields are observed in complex and hierarchical structures covering a wide range of scales that emerge and vanish on the time-scales of turbulent convective patterns. Traditionally, they are classified by size and lifetime as patterns of granulation (1 Mm, 0.2 h), mesogranulation (5-10 Mm, 5 h) and supergranulation (15-35 Mm, 24 h). Recently, the idea is emerging that meso and supergranulation are signatures of a collective interaction of granular cells \citep{delmoro07}. The continuous restructuring of surface magnetic fields by photospheric turbulent flows plays a key role in determining the topology and evolution of chromospheric and coronal magnetic fields, and may influence the triggering of eruptive solar events such as flares, coronal mass ejections or sudden disappearance of filaments \citep{roudier09}. For example,
\citet{rondi07} and \citet{roudier08} used space and ground observations to show that large-scale horizontal photospheric
flows below and around a filament influence the formation and evolution of filaments, leading to destabilization
of the coronal magnetic field.

Spectral line observations of the Sun with high spatial resolution show small-scale bright points, with widths
$ \lesssim 0.5$ Mm, located in the intergranular lanes. Bright points are associated with localized regions of strong magnetic field and correspond to magnetic flux tubes of kilogauss field strength that stand nearly vertically in the solar atmosphere \citep{solanki93}. Some bright points exhibit the proper motions of convectively driven vortical flows created at the downdrafts where the plasma follows spiral paths and returns to the solar interior after cooling down \citep{bonet08}. Recent balloon-borne Sunrise observations show $3.1 \times 10^{-3}$ vortices Mm$^{-2}$ min$^{-1}$, with a mean lifetime of 7.9 min and a standard deviation of 3.2 min \citep{bonet10}. These magnetized vortex tubes have been predicted by the theory of magnetoconvection \citep{axel96,stein98} and observed in 3D radiative/compressible MHD simulations \citep{kitiashvili10,kitiashvili12,shelyag11,rempel13}. Photospheric vortex structures with high concentration of magnetic field provide an important path for energy and momentum transfer from the convection zone into the chromosphere \citep{kitiashvili12} and may be relevant for chromospheric-coronal heating \citep{ballegooijen98,hasan08}.

A number of observational and theoretical works have applied the Lagrangian approach to trace the flow of bright points and the pattern of the magnetic network. \citet{simon88} measured the horizontal flow field on the solar surface using the technique of local correlation tracking (LCT) on a 28 min time sequence of white-light images of the Solar Optical Universal Polarimeter instrument onboard the Spacelab2. They showed that insight into the relationship between flows and magnetic fields is acquired by calculating the flow of passive test particles, called ``corks'', that are originally distributed uniformly in the flow field. The cork paths, representing the Lagrangian tracers of the pathlines of the flow, congregate at the same locations of the magnetic network. A simplified kinematic model of convection at the solar surface was developed by \citet{simon89} to interpret observations and predict the evolution of magnetic fields. They compared the cork patterns of the simulated flow with the cork patterns of the observed magnetic network, and discussed the extent to which magnetic flux tubes can be regarded as moving passively with the large-scale photospheric flow. \citet{ballegooijen98} used the G-band images of the Swedish Vacuum Solar Telescope to show that bright points are arranged in linear structures, called ``filigrees'', located in the lanes between neighboring granule cells and to measure the motion of bright points using an object tracking technique; comparison with a 2D simulation of horizontal motions of magnetic flux elements in response to solar granulation flows shows that the observed velocities and spatial distribution of the bright points are consistent with passive advection of corks by the solar granulation flow.  \citet{delmoro07} reconstructed the 3D velocity field of a single supergranular cell from the spectrometer data of the Interferometric Bidimensional Spectrometer, and used the tracking of corks to show that a divergence structure is created by a compact region of constant up-flow close to the cell centre, and isolated regions of strong convergent down-flow are nearby or cospatial with extended clusters of bright CaII wing features forming the knots of the magnetic network. 
\citet{schmieder13} applied the coherent structure tracking (CST) algorithm to track granules and analyze large-scale photospheric flows related to an extended filament in the active region AR 11106 using the $H_{\alpha}$ image of the THEMIS telescope and the EUV imager (AIA) of Solar Dynamic Observatory. They showed that diverging flows inside the supergranules may be similar in and out of the filament channel. In addition, they identify converging flows, corresponding to accumulation of corks,  around the footpoints or
barbs located at the edges of the EUV filament. The frequent EUV brightenings suggest the occurrence of reconnections of the magnetic field lines of the flux tube with the environment at the convergence points at the edges of the EUV filament channel.

There is observational evidence of anomalous transport and turbulence in photospheric flows and magnetic fields. \citet{schrijver92} and \citet{lawrence93} applied the percolation theory to report sub-diffusion of magnetic elements in the photosphere. \citet{cadavid99} used a 70 min sequence of G-band image of the Swedish Vacuum Solar Telescope on 5-October-1995 to show that the transport of bright points near the disk center is subdiffusive due to the trapping of walkers at the stagnation points in the fractal intercellular pattern; the distribution of waiting times at the trap sites obeys a L\'evy (power-law) distribution. \citet{lawrence01} applied subsonic filtering to the data of \citet{cadavid99} to report super-diffusion and show that the spatiotemporal scaling of the bright point dynamics indeed indicates the presence of turbulence in the intergranular lanes. \citet{lawrence11} used a 32 min sequence of G-band and Ca II K-line intensity measurements of the Dunn Solar Telescope on 28-May-2009, at the disk center of size 32 Mm $\times$ 32 Mm, to investigate high-frequency fluctuations in both photosphere and chromosphere; their noise-corrected G-band spectrum for $f = 28-326$ MHz shows a power law with exponent $-1.21 \pm 0.02$, consistent with the presence of turbulent motions. Moreover, they showed that the G-band spectral power in the 25-100 MHz range is concentrated at the locations of magnetic bright points in the intergranular lanes and is highly intermittent in time, being characterized by a positive kurtosis, which implies that the fluctuations have non-Gaussian probability distribution functions (PDF) with broad tails indicative of turbulence. \citet{cadavid12} showed that the noise-filtered power spectrum of the Hinode/SOT data with a sequence of 80 min on 5-March-2007, in the internetwork near the Sun center of size 20 Mm $\times$ 20 Mm, presents a scaling range of $32 < f < 53$ MHz with characteristic power law exponents of -1.56 for G-band and of -4.00 for H-line, confirming the presence of turbulent fluctuations. By integrating the Hinode/SOT G-band spectra in the range 32-53 MHz, they identified the sites of increased G-band power with the location of G-band bright points. \citet{abramenko11} and \citet{lepreti12} used the images of New Solar Telescope of Big Bear Solar Observatory  to report super-diffusion and turbulent pair dispersion of bright points in active region plage, quiet-sun, and coronal hole.

The aim of this paper is to apply Eulerian and Lagrangian tools to detect coherent structures in images of photospheric turbulent flows. First, we apply the Eulerian  approach to compute the Q-criterion
and distinguish regions of the plasma flow dominated by the deformation and vortical coherent structures \citep{hunt88}. Next, we apply the tool of Lagrangian coherent structures (LCS) to determine the repelling and attracting material lines that organize the transport of plasma flow. The term ``Lagrangian'' refers to flows defined by the fluid motion instead of an instantaneous snapshot (Eulerian); the term ``coherent'' refers to the distinguished stability of these structures compared to other nearby material lines/surfaces. There is a growing interest to adopt the LCS approach \citep{haller00,shadden11} to improve the understanding of transport in complex flows such as planetary atmosphere \citep{sapsis09,peng12}, oceans \citep{beron-vera10,lehahn11}, and cardiovascular system \citep{shadden08,arzani12}. A number of papers have studied LCS in complex plasma flows for thermonuclear applications \citep{leoncini06,padberg07,borgogno11} and astrophysical applications \citep{rempel11,rempel12,rempel13,yeates12}. In this paper, we apply the backward finite-time Lyapunov exponents (b-FTLE) to study the attracting LCS of the photospheric turbulence.
We show that, like the ``corks'', the b-FTLE is able to trace the patterns of the magnetic network. Moreover, it gives
a quantitative measure of the local rate of contraction/expansion of the photospheric flow.

\section{The horizontal velocity data}

Our study is based on a 12 h high-resolution sequence of velocity images (cadence $\sim121 s$) of photospheric flow derived from the Hinode-SOT  magnetograms, from 12-Dec-2006 14:24 UT to 13-Dec-2006 02:24 UT, by selecting a unipolar plage area of size 12.4 Mm $\times$ 12.4 Mm ($\sim17''\times 17''$) near the active region AR 10930 (Yeates et al. 2012). The horizontal velocity field is a ``proxy'' extracted from the line-of-sight magnetic field ($B_z$) using the Fourier local correlation tracking method (FLCT) (Welsch et al. 2004, 2012).

It is worth pointing out that there are uncertainties in the correlation tracking that
do not fully resolve the velocity fields in either space or time. The detailed
description of the data reduction procedure is given in \citet{welsch12} and \citet{yeates12}.
The FLCT method involves a number of optimum parameters which are determined by an autocorrelation
analysis in order to maximize frame-to-frame correlations and ensure robustness in the
velocity estimate. In view of the reduction of noise by averaging in the tracking procedure,
the actual velocity patterns of the photospheric flow might be more complicated than the ``proxy''.

\citet{yeates12} showed that the build-up of magnetic gradients in solar corona can be
inferred directly from the photospheric velocity data, without recourse to magnetic field
extrapolation. They established the correspondence of the network of quasi-separatrix layers
\citep{demoulin96} to the repelling LCS in the photospheric velocity. In this 
paper, we extend the analysis of \citet{yeates12} to establish the correspondence of the 
network of high magnetic flux concentration to the attracting LCS in the photospheric velocity.

\section{Eulerian coherent structures in the photosphere}

Eulerian coherent structures can be extracted from the velocity field by computing the Q-criterion, also known as the 
Okubo-Weiss parameter in 2D turbulence \citep{hunt88}, 

\begin{equation}
Q = \frac{1}{2}[|\Omega|^2 - |S|^2],
\label{eq q}
\end{equation}
where the Frobenius matrix norm is adopted and  $S$ and $\Omega$ are defined by the decompositions of the gradient tensor of the velocity field 

\begin{equation}
\nabla u = S + \Omega,
\label{eq u} 
\end{equation}
and 

\begin{equation}
S = \frac{1}{2}[\nabla u + (\nabla u)^{\top}], \quad \Omega = \frac{1}{2}[\nabla u - (\nabla u)^{\top}],
\end{equation}
where $\top$ denotes the transpose of a tensor. The strain-rate (or deformation) tensor $S$ is the symmetric part of $\nabla u$; its eigenvectors form an orthonormal set along the principle directions of fluid element deformation caused by tension, and its eigenvalues are real quantities that measure the rates of deformation in the corresponding directions. The vorticity tensor $\Omega$ is the anti-symmetric part of $\nabla u$. For example, the deformation calculated using the Ulysses data shows that the magnetic tension causes local stretching of solar wind plasma elements mostly along the direction of the average magnetic field, whereas the solar wind vorticity tends to become perpendicular to the mean radial direction at large heliospheric distances \citep{polygiannakis96}.

Snapshot images of Eulerian coherent structures of the horizontal velocity field detected by Hinode in a plage region near AR 10930 \citep{yeates12} are shown in Figure \ref{fig euler} for four instances: 12-Dec-2006 14:24 UT, 12-Dec-2006 18:24 UT, 12-Dec-2006 22:24 UT, 13-Dec-2006 02:24 UT. The first column shows magnetograms; the second column depicts the streamlines computed using the method of line integral convolution, which displays the integral curves of $(u_x, u_y)$ in different tones of orange; some vortex coherent structures (patches) can be visualized in these plots. The third column depicts the $(x, y)$ plots of the deformation Eulerian coherent structures, which are identified by the localized patches with high values of $|S|^2$. The fourth column depicts the $(x, y)$ plots of the vortical Eulerian coherent structures, which are identified by the localized patches with high values of $|\Omega|^2$. The average spatial size and lifetime of these vortices are similar to the magnetic bright points observed by \citet{bonet08,bonet10}.  The fifth column depicts the $(x, y)$ plots of the Q-criterion given by Eq. (1), which partitions the turbulent flow into the deformation-dominated  ($Q <  0$, blue patches) and vorticity-dominated ($Q > 0$, red patches) regions.

The intensity of a turbulent flow is characterized by the magnitude of the root-mean-squared velocity fluctuations, deformation, and vorticity \citep{steinberg09}. The probability distribution functions (PDF) of Eulerian coherent structures, associated with Figure \ref{fig euler}, are shown in Figure \ref{fig pdfeuler}. Some trends are evident. Figures \ref{fig pdfeuler}(a) and \ref{fig pdfeuler}(b) show that there are higher (lower) statistics for the low (high) values of the magnitudes of deformation and vorticity. Figure \ref{fig euler}(c) shows that for all four instances the Q-criterion is distributed amongst negative and positive values, with non-Gaussian PDF. Note that the PDF of the Q-criterion shows a strong asymmetry skewed towards negative values, which indicates that in the plage region under study the photospheric turbulent flow is dominated by the strain-rate field. This situation is similar to a laboratory experiment of the turbulence-flame interaction, where the flame surface straining is dominated by the deformation whereas the role of the vorticity is to curve the flame surface creating wrinkles \citep{steinberg09}.

The mean values of Eulerian coherent structures for four instances are summarized in Table 1.
It confirms the features of PDF seen in Figure \ref{fig pdfeuler}, i.e., in the plage region under study, on average: (i) the deformation dominates over the vorticity, (ii) the Q-criterion shows negative values implying the dominance of the deformation coherent structures over the vortical coherent structures.

\begin{table}[!h]
\caption{Mean values of the Eulerian coherent structures}
\begin{tabular}{|c|c|c|c|}
\cline{2-4} 
\multicolumn{1}{c|}{} & $\left\langle |S|^{2}\right\rangle $ & $\left\langle |\Omega|^{2}\right\rangle $ & $\left\langle Q \right\rangle$\tabularnewline
\hline 
12-Dec-2006 14:24 UT & 5.0453e-08 & 1.8139e-08 & -1.6157e-08\tabularnewline
\hline 
12-Dec-2006 18:24 UT & 7.3680e-08 & 2.4142e-08 & -2.4769e-08\tabularnewline
\hline 
12-Dec-2006 22:24 UT & 8.0154e-08 & 2.2815e-08 & -2.8670e-08\tabularnewline
\hline 
13-Dec-2006 02:24 UT & 4.9614e-08 & 1.4840e-08 & -1.7387e-08\tabularnewline
\hline 
\end{tabular}
\end{table}

Although the deformation and vortical coherent structures can be extracted by the Eulerian technique using the Q-criterion, it relies on 
instantaneous snapshots of the velocity field. In the next section we show that the transport barriers in the turbulent flow can be detected with greater precision by the technique of LCS.

\section{Lagrangian coherent structures in the photosphere}

Analogous to ``corks'' discussed in Section 1, the finite-time Lyapunov exponent (FTLE) can be computed by advecting a dense grid of tracer particles over the domain of interest. Consider a passive particle advected by the velocity field $\mbox{\bf u}(\mbox{\bf r}, t)$ from an initial time $t_0$. We use the horizontal velocity field derived from the Hinode SOT data for the plage region of Figure \ref{fig euler} to solve the particle advection equation 

\begin{equation}
\frac{d\mbox{\bf r}}{dt} = \mbox{\bf u}(\mbox{\bf r}, t), \quad \mbox{\bf r}(t_0) = \mbox{\bf r}_0, 
\label{eq dudt}
\end{equation}
over a grid of initial positions $\mbox{\bf r}_0$ until the final positions $\mbox{\bf r}(t_0 + \tau)$ are reached after a finite integration time 
$\tau$. The particle trajectories are obtained by a fourth-order Runge-Kutta integrator 
with cubic spline interpolation in both space and time. The finite-time Lyapunov exponents of the particle trajectories for a 2D flow are calculated at each initial position $\mbox{\bf r}_0$ as \citep{shadden05,shadden11,rempel11,rempel12,rempel13,yeates12}

\begin{equation}
\sigma^{t_0+\tau}_i (\mbox{\bf r}_0) = \frac{1}{|\tau|}\ln \sqrt{\lambda_i}, \quad   i = 1,2,
\label{eq lambda}
\end{equation}
where $\lambda_i$ ($\lambda_1 > \lambda_2$) are the eigenvalues of the finite-time right Cauchy-Green deformation tensor $\Delta = J^{\top} J$,  
$J = d\phi^{t_0+\tau}_{t_0} (\mbox{\bf  r})/d\mbox{\bf r}$ is the deformation gradient, $\top$ denotes the transpose
and $\phi^{t_0+\tau}_{t_0}:\mbox{\bf  r}(t_0)\rightarrow \mbox{\bf  r}(t_0+\tau)$ is the flow map for Eq. $(\ref{eq dudt})$. 
In forward time $(\tau>0$), the maximum FTLE, $\sigma_1$, gives the finite-time average of the maximum rate of either divergence (if $\sigma_1>0$) or convergence (if $\sigma_1<0$) between the trajectories of a fiducial particle at 
$\mbox{\bf r}$ and its neighboring particles. The maximum stretching is found when the neighboring particle $\mbox{\bf s}$ is such that $\xi_0 =\mbox{\bf r}_0  - \mbox{\bf s}$ is initially aligned with the eigenvector of $\Delta$ associated with $\lambda_1>0$. If two points are initially separated by a small distance $|\xi_0|$ at time $t_0$, then their maximum separation at a later time $t_1$ will be $|\xi_1| \sim \exp[\sigma_1^{t_1}(\mbox{\bf r}_0)|t_1-t_0|]|\xi_0|$. $\sigma_2$ provides information about stretching/contraction in another direction and can be useful to interpret the local dynamics of the fluid. Local minima in the maximum forward finite-time Lyapunov exponent (f-FTLE) provides a way to detect the position of the center of vortices in the velocity field since vortices may be viewed as material tubes of low particle dispersion.
 In backward time $(\tau<0$), $\sigma_1>0$ represents regions of the flow with convergence and $\sigma_1<0$, 
regions with divergence of nearby trajectories. 

Advecting a particle forward in time reveals the repelling LCS in the f-FTLE field, which are the source of stretching in the flow, whereas advecting a particle backward in time reveals the attracting LCS in the b-FTLE field along which particles congregate to form the observable patterns \citep{sapsis09}. 
The properties of an LCS depend on the choice of the integration time $\tau$, which should be chosen long enough for dominant features to be revealed, yet short enough for the FTLE to be representative of the transient dynamics of interest.

Three examples of repelling LCS for the phostospheric turbulent flow in the plage region under study are given in Figures \ref{fig lcs}(a)-(c), which show the maximum f-FTLE $\sigma_1$ computed by solving the particle advection equation (4) from the initial time $t_0$ = 12-Dec-2006 14:24 UT to the final time $t_0 + \tau$, with $\tau = +4$h (Figure \ref{fig lcs}(a)), $+8$h (Figure \ref{fig lcs}(b)), and $+12$h (Figure \ref{fig lcs}(c)), respectively. Thin ridges of large (positive) f-FTLE in Figures \ref{fig lcs}(a)-(c) represent the locally strongest repelling material lines in the photospheric flow, which confirm the results of \citet{yeates12} using $\tau= +6$h and +12h. As we increase integration time we notice some structures persist and become more sharply defined. These structures can be considered most influential over time. The structures that fail to persist as integration time is increased can be considered to be due to more transient flow features.  

Three examples of attracting LCS for the photospheric turbulent flow in the plage region under study are given in Figures \ref{fig lcs}(d)-(f), which show the maximum b-FTLE $\sigma_1$ computed by solving the particle advection equation (4) from the initial time $t_0$ = 13-Dec-2006 02:24 UT to the final time $t_0 + \tau$, with $\tau = -4$h (Figure \ref{fig lcs}(d)), $-8$h (Figure \ref{fig lcs}(e)), and $-12$h (Figure \ref{fig lcs}(f)), respectively. Thin ridges of large (positive) b-FTLE in Figures \ref{fig lcs}(d)-(f) represent the locally strongest attracting material lines in the photospheric flow.

\begin{figure*}
\begin{center}
 \includegraphics[width=0.85\columnwidth]{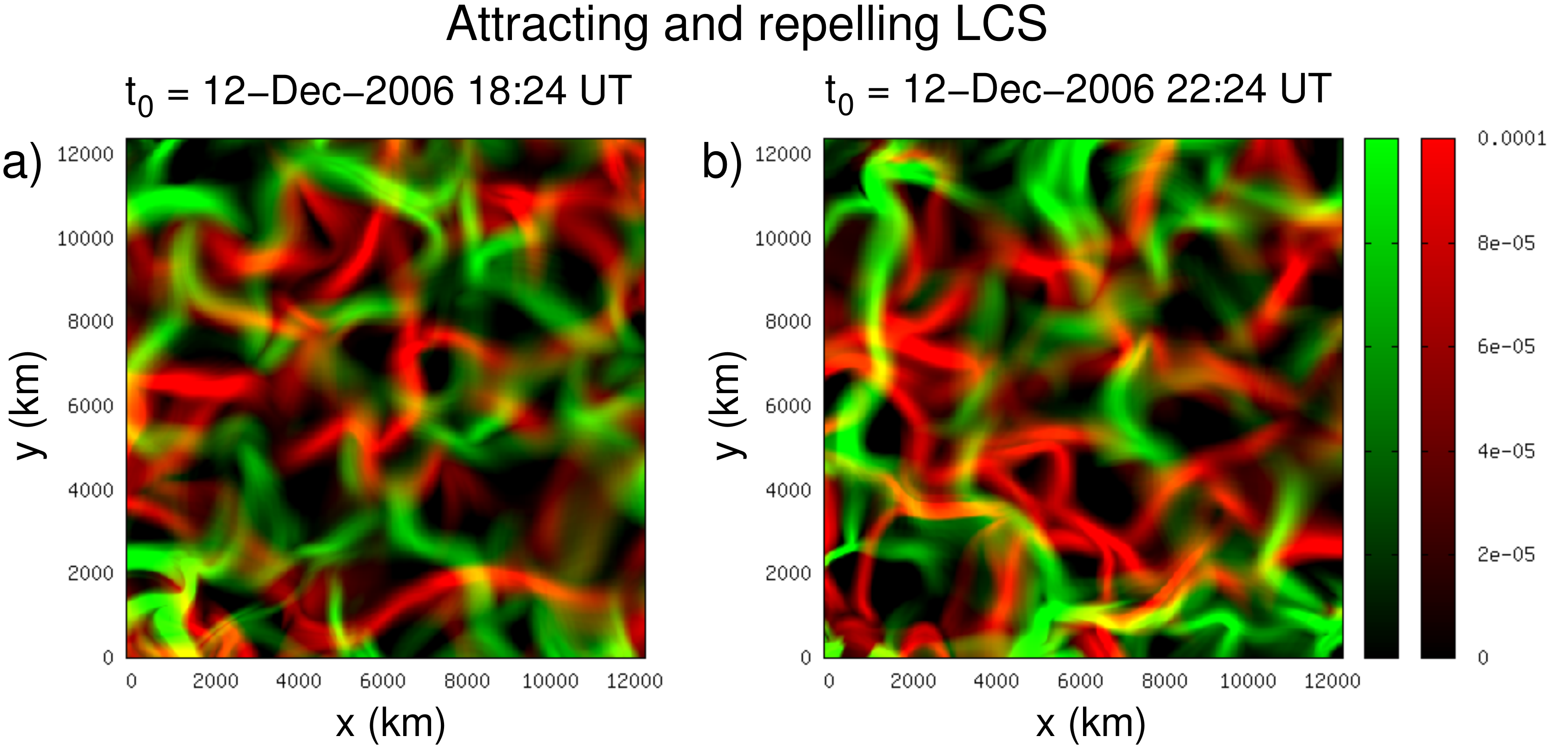}
\end{center}
 \caption{\label{fig lcs2} {\bf Attracting and repelling Lagrangian coherent structures given respectively by the b-FTLE ($s^{-1}$) (red) and f-FTLE ($s^{-1}$) (green)} at {\bf (a)} 12-Dec-2006 18:24 UT, {\bf (b)} 12-Dec-2006 22:24 UT.}
 \end{figure*}

In order to illustrate the temporal evolution of LCS, we compute the forward- and backward-time FTLE of the photospheric velocity field at $t_0$ = 12-Dec-2006 18:24 UT and $t_0$ = 12-Dec-2006 22:24 UT, using the same integration time $|\tau| = 4$h. Figure \ref{fig lcs2} displays the repelling (green) and attracting (red) LCS for two $t_0$, which form the ``Lagrangian skeletons'' of the photospheric turbulence.

Attracting and repelling LCS act as barriers to particle transport. A material line is a smooth curve of fluid particles advected by the velocity field \citep{haller01}. These attracting and repelling material lines are the analogous of stable and unstable manifolds of time-independent fields \citep{shadden05,shadden11}. Numerical studies of 2D flows have elucidated the role of material lines \citep{haller00,miranda13}. Consider a steady flow, where the velocity field does not change with time. In the presence of counter-rotating vortices, hyperbolic (saddle) points are expected to be found. The trajectories of passive scalars follow the velocity vectors in the vicinity of the hyperbolic point. Thus, particles lying on the stable manifold are attracted to the saddle point in the forward-time dynamics and trajectories on the unstable manifold converge to the saddle point in the backward-time dynamics \citep{rempel12}. Two particles are said to straddle a manifold if the line segment connecting them crosses the manifold. The maximum FTLE typically has high values on the stable manifold in forward-time, since nearby trajectories straddling the manifold will experience exponential divergence when they approach the saddle point. Similarly, the FTLE field exhibits a local maximizing curve (ridge)
along the unstable manifold in backward-time dynamics, since trajectories straddling the unstable manifold diverge exponentially when they approach the saddle point in reversed-time. Thus, ridges in the forward-time FTLE field mark the stable manifolds of hyperbolic points and ridges in the backward-time FTLE field mark the unstable manifolds. Analogously, for a time-dependent velocity field, regions of maximum material stretching generate ridges in the FTLE field. Thus, the repelling material lines (finite-time stable manifolds) produce ridges in the maximum FTLE field in the forward-time system and attracting material lines (finite-time unstable manifolds) produce ridges in the backward-time system, as seen in Figures \ref{fig lcs} and \ref{fig lcs2}.

\section{Discussion and Conclusions}

In this paper, we showed that Eulerian coherent structures and Lagrangian coherent structures give very different information of the dynamics and structure of astrophysical plasma flows. Eulerian coherent structures give instantaneous information of plasma dynamics and structure at a given time, whereas Lagrangian coherent structures account for the integrated effect of plasma dynamics and structure in a finite-time interval. Arguably, instantaneous quantities may be considered inappropriate to understand inherently transient phenomenon since such quantities may not properly convey the integrated behavior of constantly changing flow. LCS typically account for such integrated behavior more naturally by considering the integrated fluid motion to reveal organizing flow features.

\begin{figure*}
\begin{center}
 \includegraphics[width=1\columnwidth]{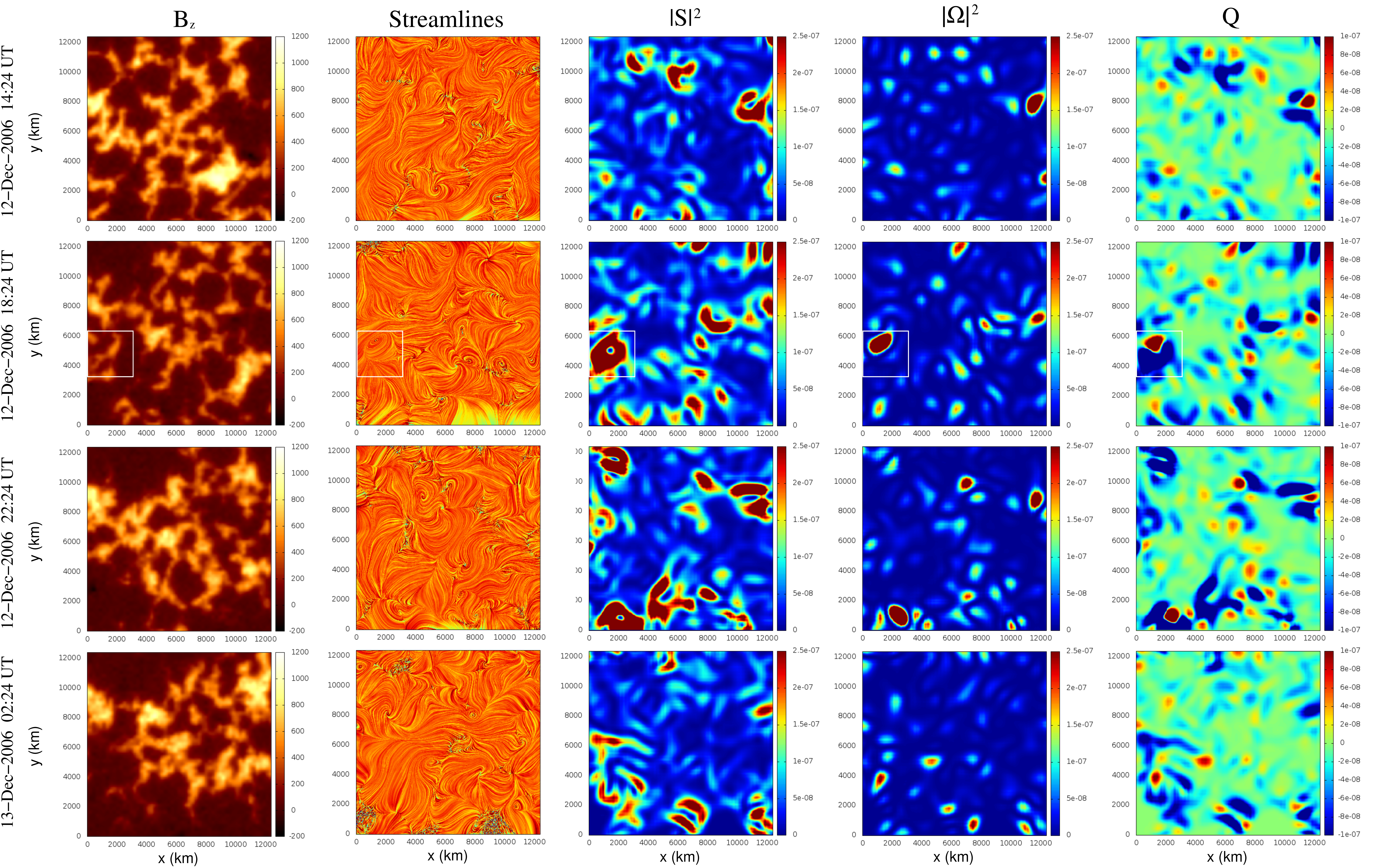}
\end{center}
 \caption{\label{fig euler} {\bf
Images of the magnetic field $B_z(G)$, streamlines, deformation (S) and vortical (Omega) Eulerian coherent structures, and Q-criterion} 
at: 12-Dec-2006 14:24 UT, 12-Dec-2006 18:24 UT, 12-Dec-2006 22:24 UT, 13-Dec-2006 02:24 UT (from top to bottom).}
 \end{figure*}

\begin{figure}[!h]
\begin{center}
 \includegraphics[width=0.5\columnwidth]{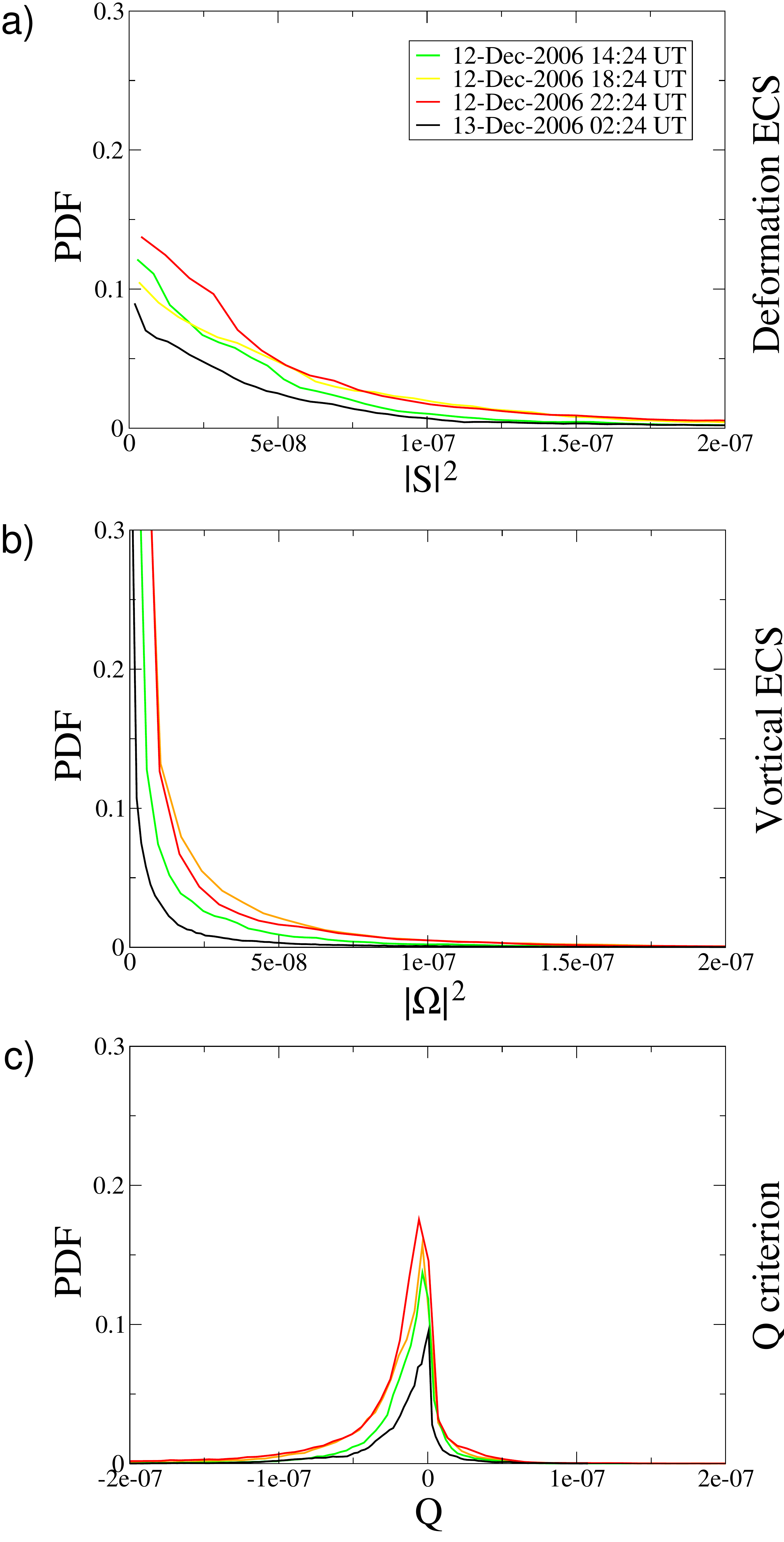}
\end{center}
 \caption{\label{fig pdfeuler} {\bf PDF of Eulerian coherent structures (ECS) as a function of: (a)} $|S|^2$,
{\bf (b)} $|\Omega|^2$, {\bf (c)} Q-criterion; at 12-Dec-2006 14:24 UT (green),
12-Dec-2006 18:24 UT (yellow), 12-Dec-2006 22:24 UT (red),
13-Dec-2006 02:24 UT (black).}
 \end{figure}

\begin{figure}
\begin{center}
\includegraphics[width=0.85\columnwidth]{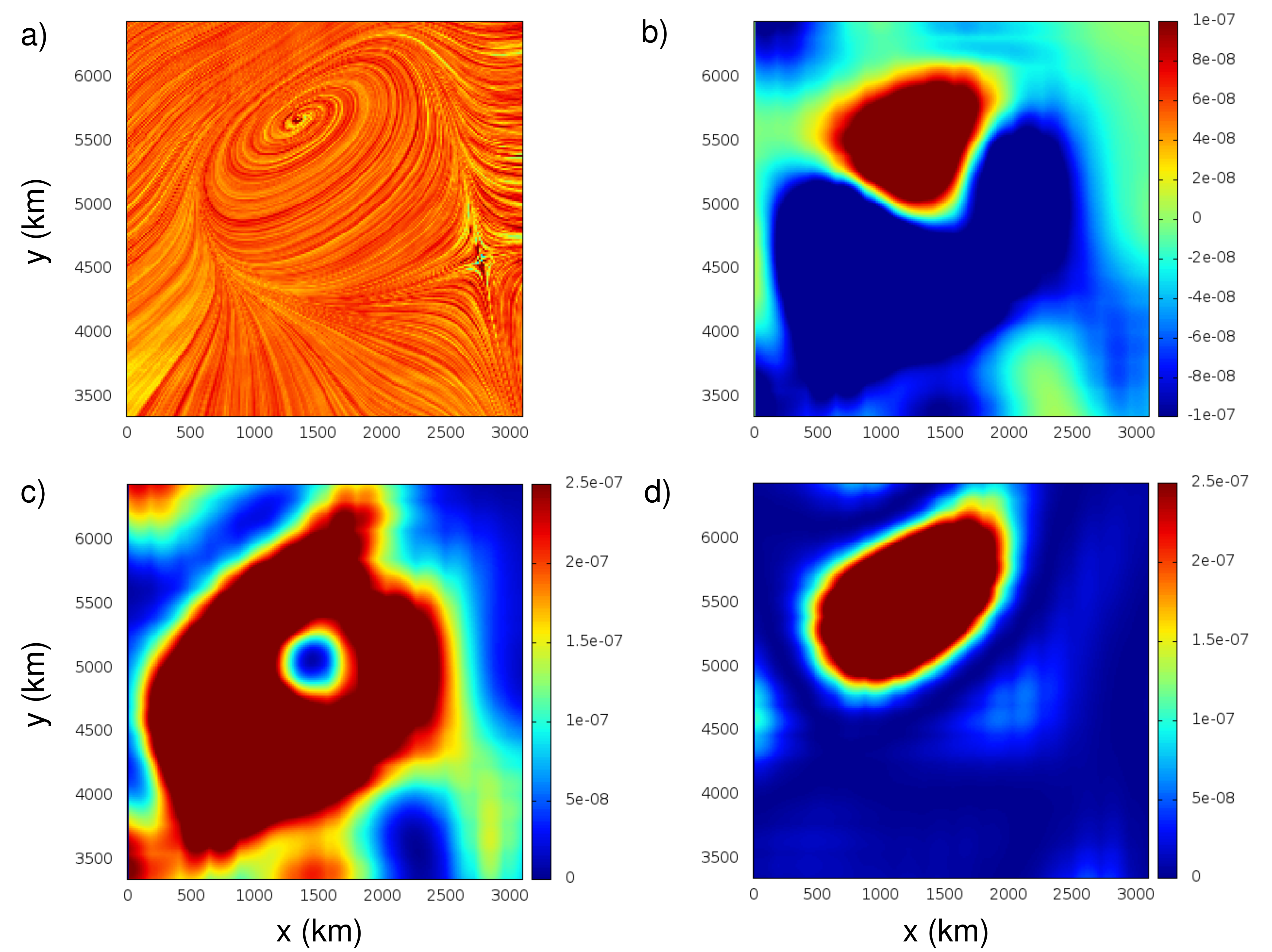}
\end{center}
 \caption{{\bf Enlargement of the rectangle region indicated at Figure \ref{fig euler} at 12-Dec-2006 18:24 UT: }
{\bf (a)} streamlines, {\bf (b)} Q-criterion, {\bf (c)} $|\Omega|^2$, {\bf (d)} $|S|^2$.}
\label{fig zoomeuler2d}
 \end{figure}

\begin{figure}
\begin{center}
\includegraphics[width=0.5\columnwidth]{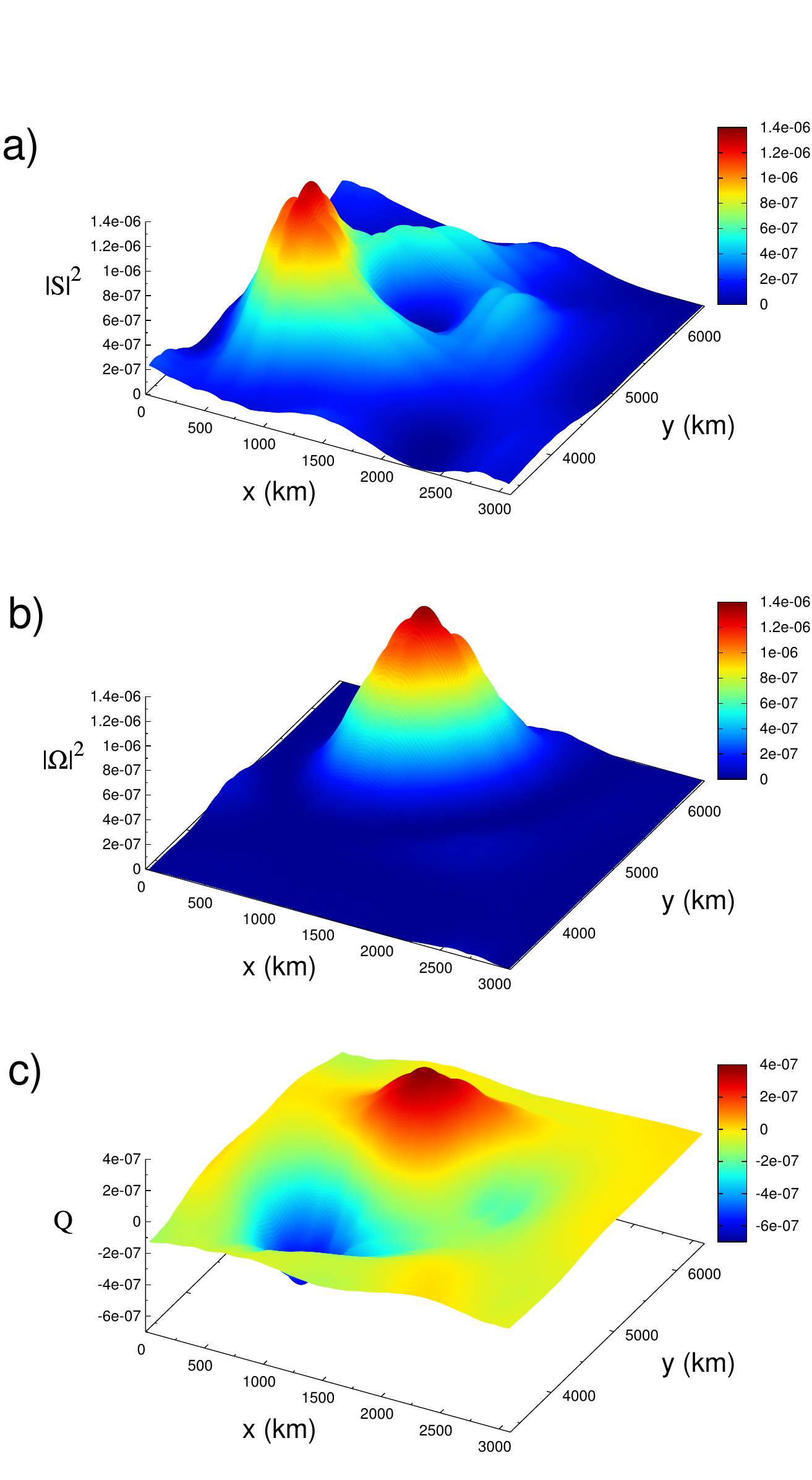}
\end{center}
 \caption{{\bf 3D plots of Eulerian coherent structures as a function of $(x, y)$, corresponding to Figure \ref{fig zoomeuler2d}.} {\bf (a)} $|S|^2$,  {\bf (b)} $|\Omega|^2$ and {\bf (c)} Q-criterion. }
\label{fig zoomeuler3d}
 \end{figure}

In the context of photospheric turbulence, we showed in Figure \ref{fig euler} that the Eulerian analysis is able to differentiate two types of coherent structures arising from either deformation or vorticity. An enlargement of the rectangle region indicated at Figure \ref{fig euler} at 12-Dec-2006 18:24 UT is depicted in Figure \ref{fig zoomeuler2d}. The pattern of streamlines in Figure \ref{fig zoomeuler2d}(a) identifies the localized region of a vortex in the photospheric flow. Figures \ref{fig zoomeuler2d}(c) and \ref{fig zoomeuler2d}(d) show that this vortex region is characterized by high values of the modulus of both deformation and vorticity. Figure \ref{fig zoomeuler2d}(b) shows that the Q-criterion can differentiate clearly two subregions: one being the deformation Eulerian coherent structures (blue subregion) and the other being the vortical Eulerian coherent structures (red subregion). For the sake of clarity, a corresponding 3D view of Figure \ref{fig zoomeuler2d} is presented in Figure \ref{fig zoomeuler3d}. Our Eulerian analysis shows that, on average, in the plage region the deformation Eulerian coherent structures dominate over the vortical Eulerian coherent structures in the turbulent photospheric flows, as seen in Figure \ref{fig pdfeuler}.

\begin{figure*}
\begin{center}
 \includegraphics[width=1.\columnwidth]{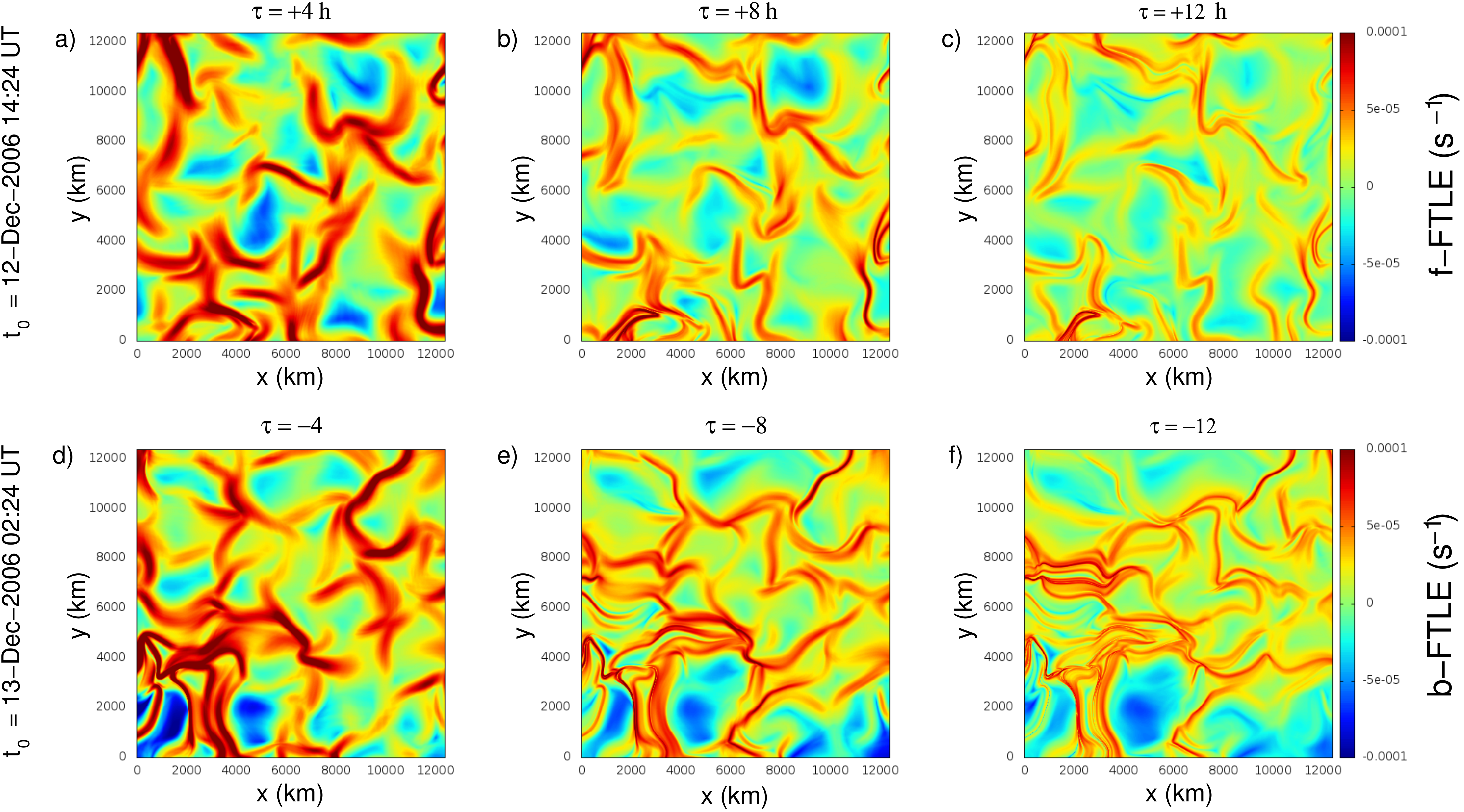}
\end{center}
 \caption{\label{fig lcs} {\bf Lagrangian coherent structures.} Upper panel: f-FTLE for $t_0$ at 12-Dec-2006 14:24 UT with {\bf (a)} $\tau = +4$ hr, {\bf (b)} $\tau = +8$ hr, {\bf (c)} $\tau = +12$ hr. Bottom panel: b-FTLE for $t_0$ at 13-Dec-2006 02:24 UT with {\bf (d)} $\tau = -4$ hr, {\bf (e)} $\tau = -8$ hr, {\bf (f)} $\tau = -12$ hr. }
 \end{figure*}

\begin{figure}
\begin{center}
\includegraphics[width=0.5\columnwidth]{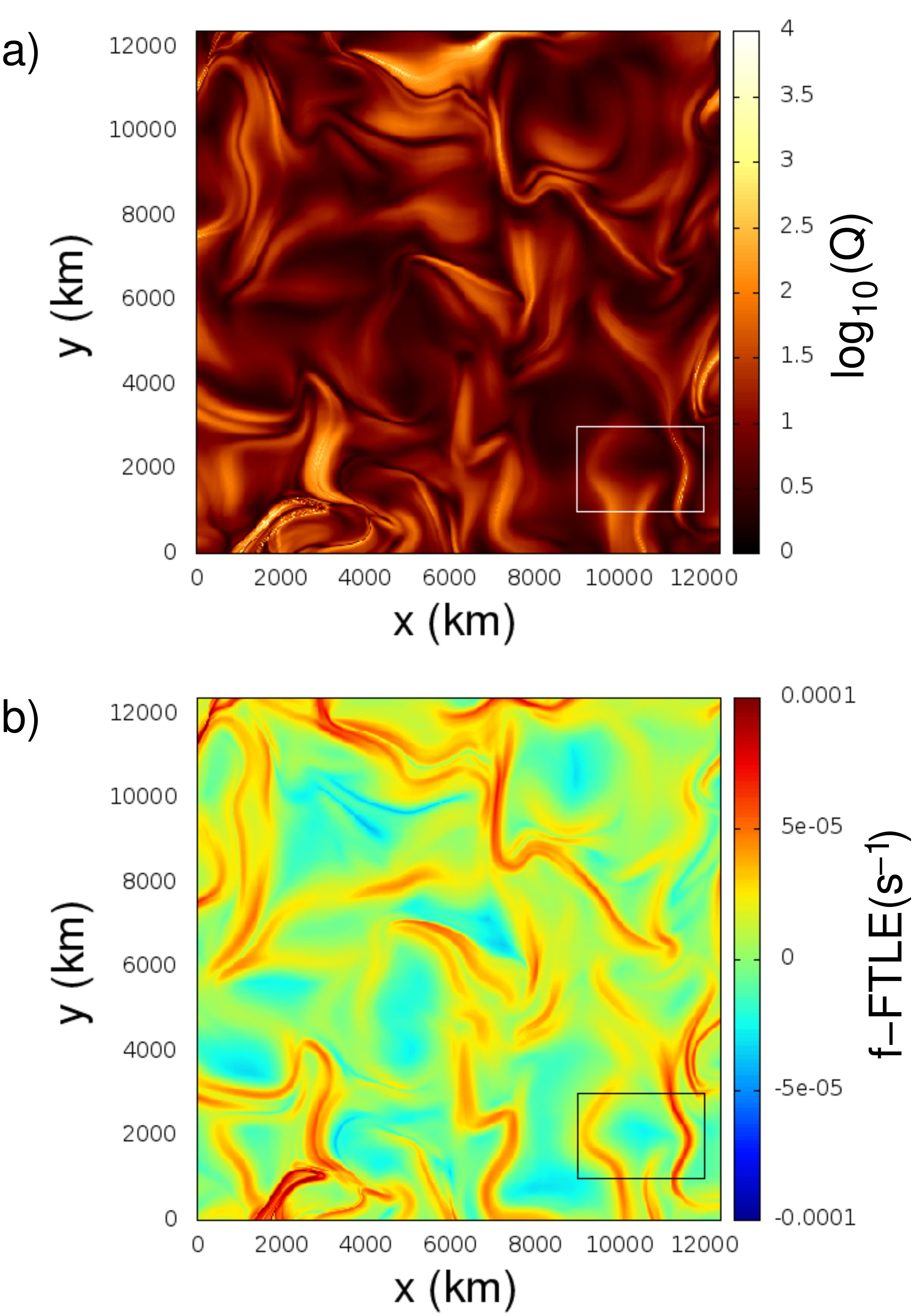}
\end{center}
 \caption{{\bf Comparison of the squashing Q-factor and repelling LCS.} {\bf (a)} Intensity plot of $\log_{10}(Q)$; {\bf (b)} the f-FTLE for $t_0$ = 12-Dec-2006 14:24 UT and $\tau$ = +12 h.}
\label{fig squash2d}
 \end{figure}

\begin{figure}
\begin{center}
\includegraphics[width=0.5\columnwidth]{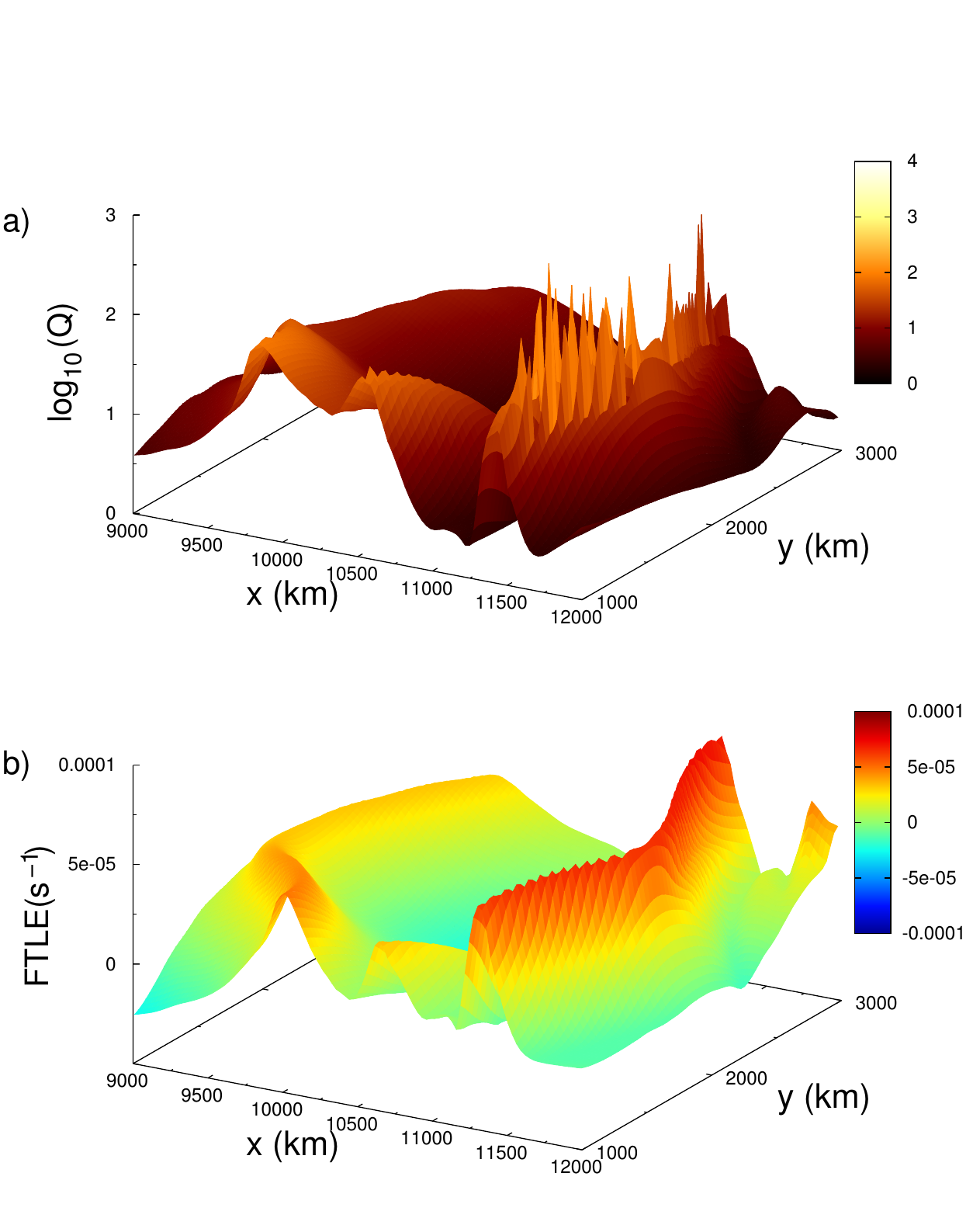}
\end{center}
 \caption{{\bf 3D plots of the squashing Q-factor and repelling LCS.} {\bf (a)} $\log_{10}(Q)$; {\bf (b)} the f-FTLE for $t_0$ = 12-Dec-2006 14:24 UT and 
$\tau$ = +12 h, corresponding to the rectangle regions indicated in Fig. \ref{fig squash2d}.}
\label{fig squash3d}
 \end{figure}

The Lagrangian analysis is able to provide further information of magnetic and velocity fields of the photospheric turbulence missing in the Eulerian analysis. Yeates et al. (2012) demonstrated that it is possible to make the link between the squashing Q-factor and the maximum Lyapunov exponent 
$\sigma_1$ of the repelling Lagrangian coherent structures, since both quantities are similar measures of the local rate of stretching at a given point and defined by the norm of the Cauchy-Green tensor of the deformation imposed by the field-line mapping. Note, however, that there are some differences between these two quantities. While the squashing Q-factor uses the Frobenius norm of the tensor, $\sigma$ uses the spectral norm; while the squashing Q-factor is dimensionless, $\sigma$ has units of inverse time and includes the logarithm in its definition in Eq. (5). The close similarity of these two quantities is shown in Figure \ref{fig squash2d}, where the squashing Q-factor calculated for the same region of the photosphere under study is shown in Figure \ref{fig squash2d}(a) and the corresponding repelling Lagrangian coherent structures are shown in Figure \ref{fig squash2d}(b). Both Figures \ref{fig squash2d}(a) and \ref{fig squash2d}(b) are computed for $t_0$ = 12-Dec-2006 14:24 UT and $\tau$ = +12 h, the same as Figure \ref{fig lcs}(c). Figure \ref{fig squash2d}(a) adopts the logarithmic scaling since certain trajectories typically become exponentially separated in time, which shows the emergence of thin ridges of high values of the squashing Q-factor representing the quasi-separatrix layers. These ridges are interspersed by regions of low values of the squashing Q-factor. Figure \ref{fig squash2d}(b) shows the emergence of thin ridges of high positive values of $\sigma_1$, representing the locally strongest repelling material surfaces in the photospheric flow. These ridges are interspersed by regions of negative $\sigma_1$ that represent converging trajectories of the plasma transport. Evidently, the ridges of the quasi-separatrix layers in Figure \ref{fig squash2d}(a) are co-spatial with the ridges of the photospheric flow in Figure \ref{fig squash2d}(b). For the sake of clarity, a 3D view of the rectangle regions of Figure \ref{fig squash2d} is presented in Figure \ref{fig squash3d}. 
Note that the spiky features seen in Figure \ref{fig squash3d}(a) may be related to numerical issues which can be smoothed by applying a refined numerical procedure \citep{partiat12}.

\begin{figure}
\begin{center}
\includegraphics[width=0.5\columnwidth]{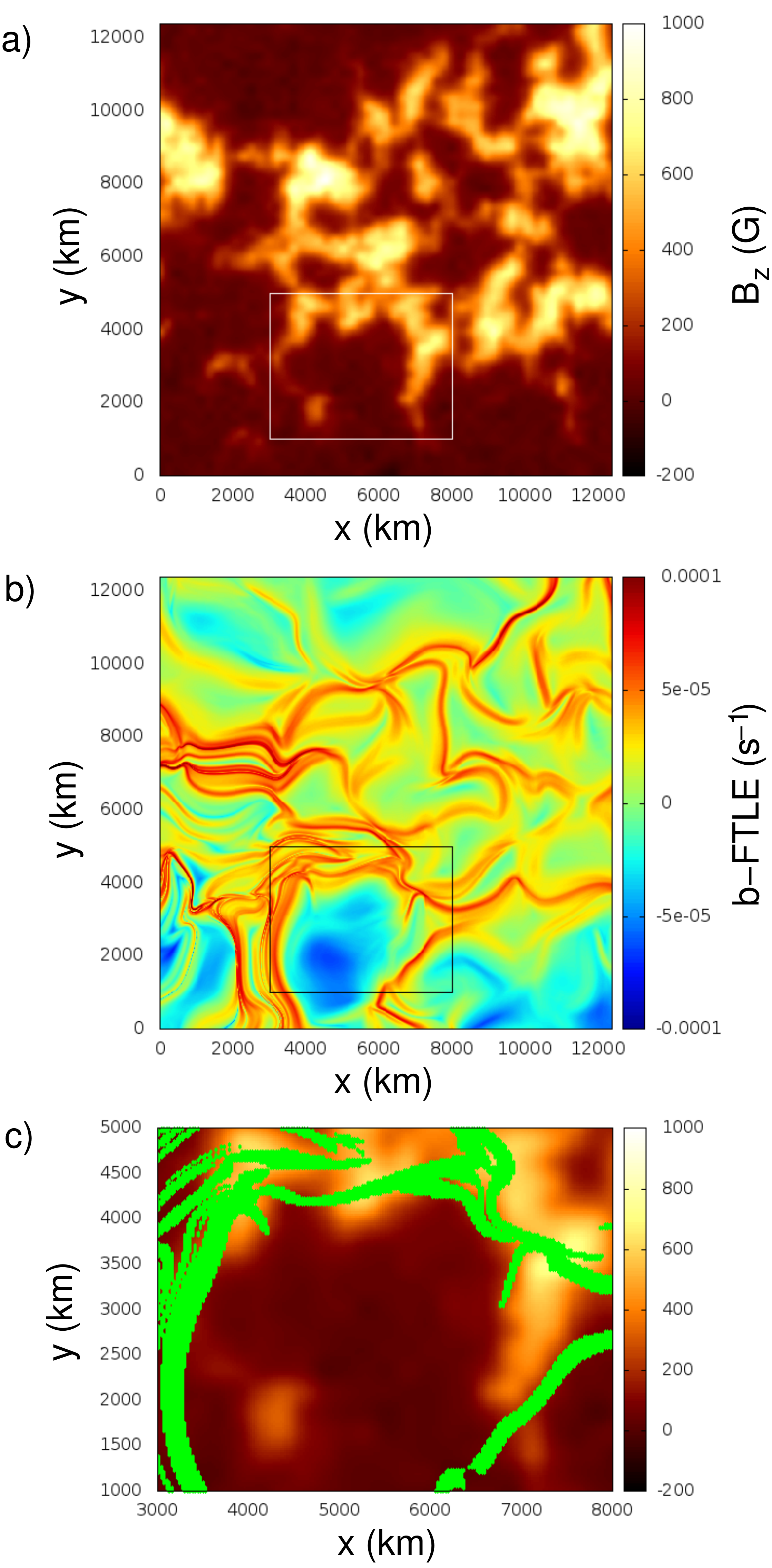}
\end{center}
 \caption{{\bf Comparison of the magnetic field and attracting LCS.} {\bf (a)} The line-of-sight magnetic field B at $t$ = 13-Dec-2006 02:24 UT and {\bf (b)} the b-FTLE for $t_0$ = 13-Dec-2006 02:24 UT and $\tau$ = -12 h. A superposition of the rectangle regions of (a) and (b) (green) is shown in {\bf (c)}, where the thresholded $\sigma_1>4\times10^{-5}$ is applied to b-FTLE.}
\label{fig blcs2d}
 \end{figure}
\begin{figure}
\begin{center}
\includegraphics[width=0.5\columnwidth]{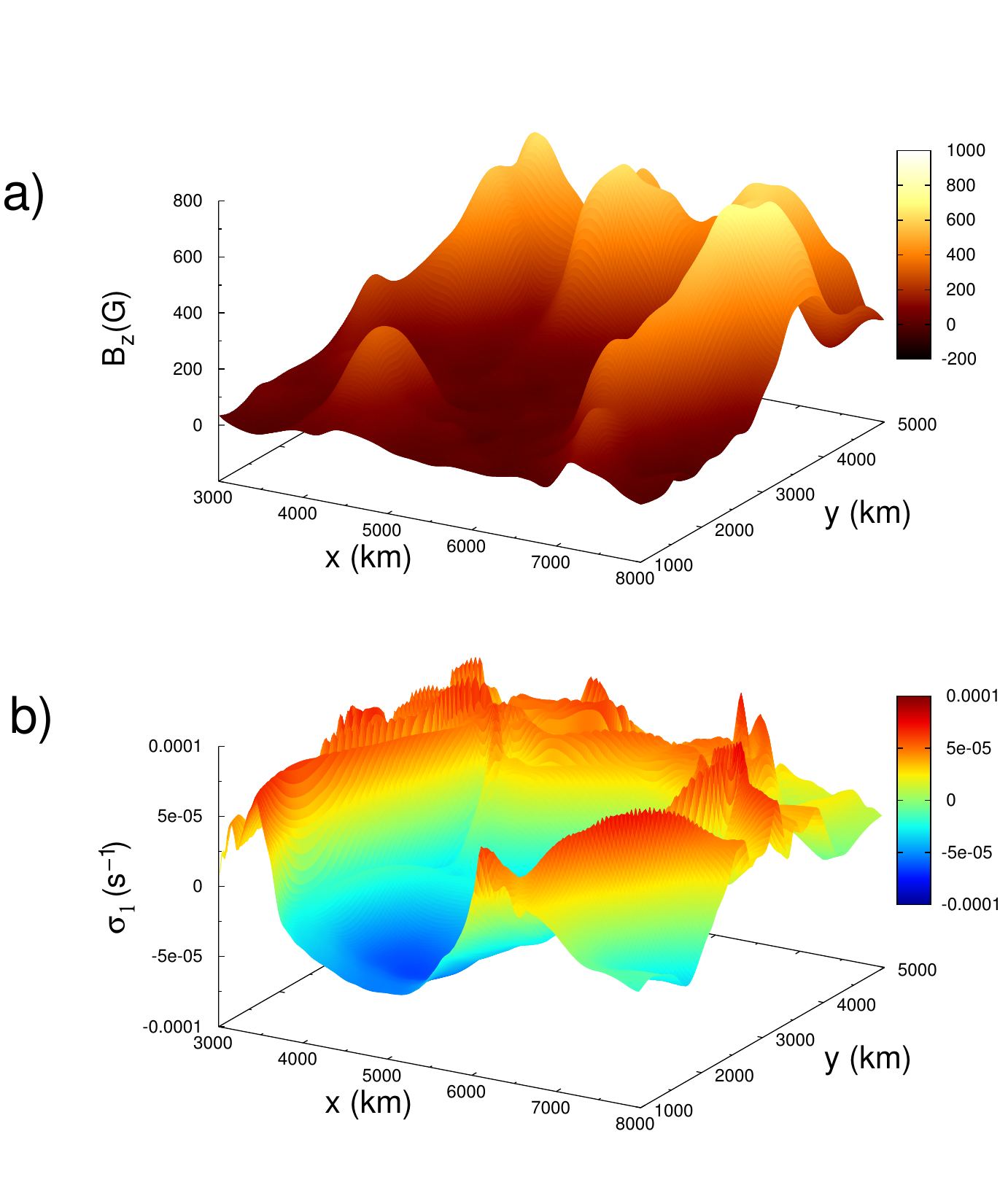}
\end{center}
 \caption{{\bf 3D plots showing the comparison of the magnetic field and attracting LCS.} {\bf (a)} The line-of-sight magnetic field $B_z$ at $t$=13-Dec-2006 02:24 UT and {\bf (b)} the b-FTLE for $t_0$ = 13-Dec-2006 02:24 UT and $\tau$ = -12 h, corresponding to the rectangle regions indicated in Figs. \ref{fig blcs2d}(a)-(b).}
\label{fig blcs3d}
 \end{figure}

As mentioned in Section 1, the Lagrangian analysis based on corks can provide insight into the relationship between magnetic fields and plasma flows in the photosphere. Roudier et al. (2009) used a 48 h time sequence of the horizontal velocity data from the Solar Optical Telescope (SOT) onboard Hinode to study the interactions between granular to supergranular scales in a quiet region of the photosphere. They showed that the tree of fragmenting granules plays a crucial role in the advection of the magnetic field and in the build-up of the magnetic network. In particular, the long time sequence of their analysis shows that the trajectory of plasma flows traced by corks matches exactly the position of the highest magnetic flux concentration of the network, which lies on supergranule boundaries. Our technique of Lagrangian coherent structures is closely related to corks, but gives additional important information missing in the cork-based studies as demonstrated below. Figure \ref{fig blcs2d} shows a comparison of the line-of-sight magnetic field $B_z$ measured by Hinode/SOT at $t$ = 13-Dec-2006 02:24 UT (Figure \ref{fig blcs2d}(a), same as Figure \ref{fig euler}) and the attracting Lagrangian coherent structures computed for $t_0$ = 13-Dec-2006 02:24 UT and $\tau$ = -12 h (Figure \ref{fig blcs2d}(b), same as Figure \ref{fig lcs}(f)). It is possible to identify similar patterns in Figures \ref{fig blcs2d}(a) and \ref{fig blcs2d}(b). For example, a superposition of the rectangle regions marked in Figures \ref{fig blcs2d}(a)-(b) is shown in Figure \ref{fig blcs2d}(c) (where the thresholded
b-FTLE with $\sigma_1>4\times10^{-5}$ is plotted) that shows clearly the proximity of the trajectory of plasma flows to the magnetic field, similar to the cork results of Roudier et al. (2009). In addition, Figure \ref{fig blcs2d}(b) shows the emergence of thin ridges of high positive values of $\sigma_1$, representing the locally strongest attracting material surfaces in the photospheric flow. These ridges are interspersed by regions of negative $\sigma_1$ that represent diverging trajectories of the plasma transport. For the sake of clarity, a 3D view of the rectangle regions of Figure \ref{fig blcs2d} is presented in Figure \ref{fig blcs3d}.

In order to confirm the validity of Figures \ref{fig blcs2d} and \ref{fig blcs3d}, we performed 3D numerical
simulations of large-scale dynamos in turbulent compressible convection with uniform horizontal
shear and rotation. The dimensionless compressible magnetohydrodynamics equations are solved in a simulation box 
divided into three layers, an upper cooling layer, a convectively unstable layer, and a stable overshoot layer,
with constant gravity in the vertical direction, as described by \citet{kapyla08}.
The box has dimensions $(L_x, L_y, L_z)=(8,8,2)d$,
where $d$ is the depth of the convectively unstable layer, and shearing periodic boundary conditions
are used in the horizontal direction. In the vertical direction we use stress-free boundary conditions for the 
velocity field and vertical field conditions for the magnetic field. 
In our simulations we adopt dimensionless quantities, setting $d=\rho_0=g=\mu_0=1$, 
where $g$ is gravity, distance is in units of $d$, density in units of the initial value at the base of convective layer $\rho_0$, 
time in units of the free fall time $\sqrt{d/g}$, velocity in units of $\sqrt{dg}$ and 
magnetic field in units of $\sqrt{d{\rho_0}{\mu_0}g}$.

The physical parameters (e.g. kinematic viscosity,
resistivity, heat conductivity, rotation velocity, etc.) are chosen to ensure the onset of a large-scale 
magnetic field in a moderately turbulent velocity field. For a detailed setup, we direct the reader to run D2 in \cite{kapyla08}.
The model is solved with the PENCIL CODE\footnote{http://pencil-code.googlecode.com/}, which employes
 sixth-order finite-differences in space and third-order variable step Runge-Kutta integration in time.

\begin{figure}
\begin{center}
\includegraphics[width=0.5\columnwidth]{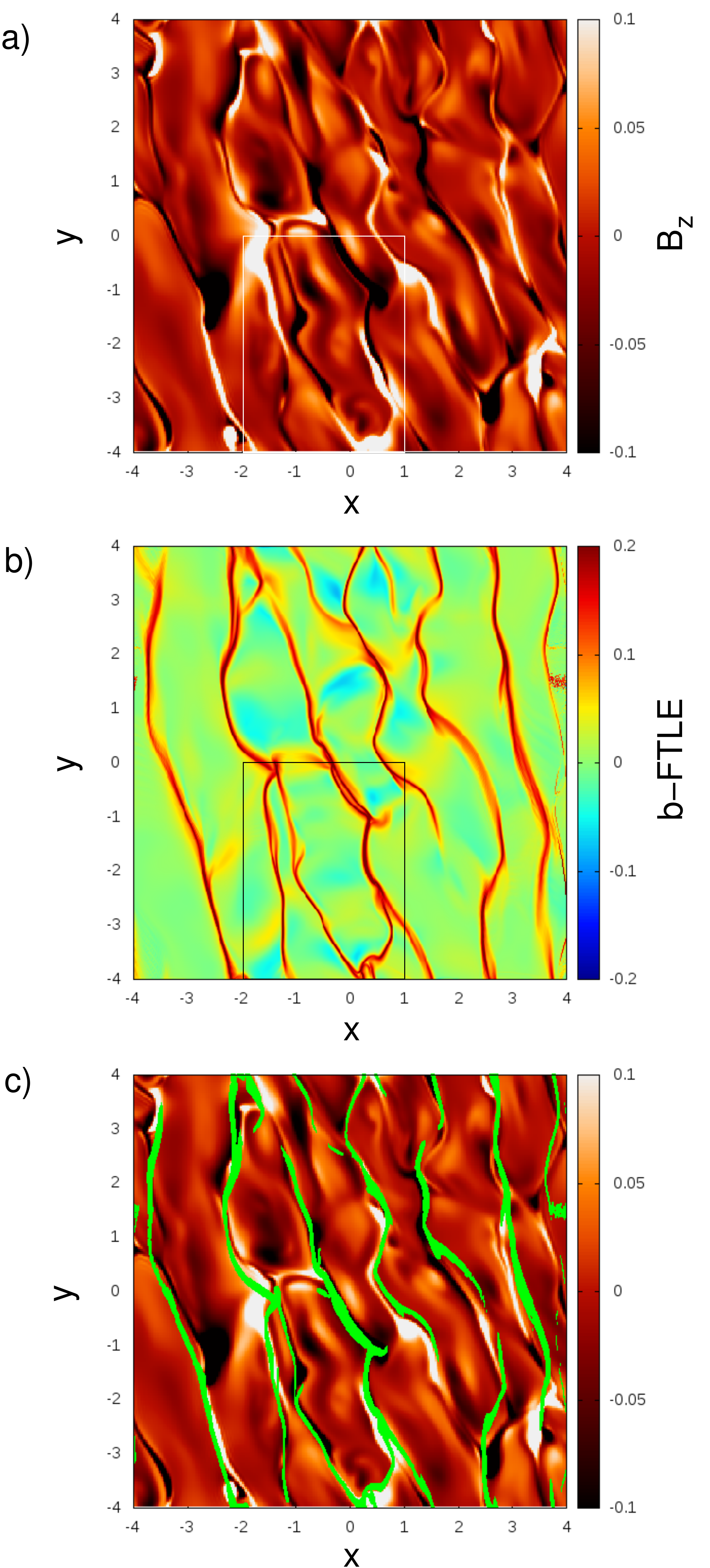}
\end{center}
 \caption{{\bf 2D plots of numerical simulation of compressible convection.} {\bf (a)} The vertical component of the magnetic field $B_z$ at time $t_0$; 
{\bf (b)} the b-FTLE computed at $t_0$ with $\tau = -10$ time units; {\bf (c)} a superposition of (a) and (b) (green), where
the threshold $\sigma_1>0.1$ is applied to b-FTLE.}
\label{fig blcs_sim2d}
 \end{figure}

\begin{figure}
\begin{center}
\includegraphics[width=0.5\columnwidth]{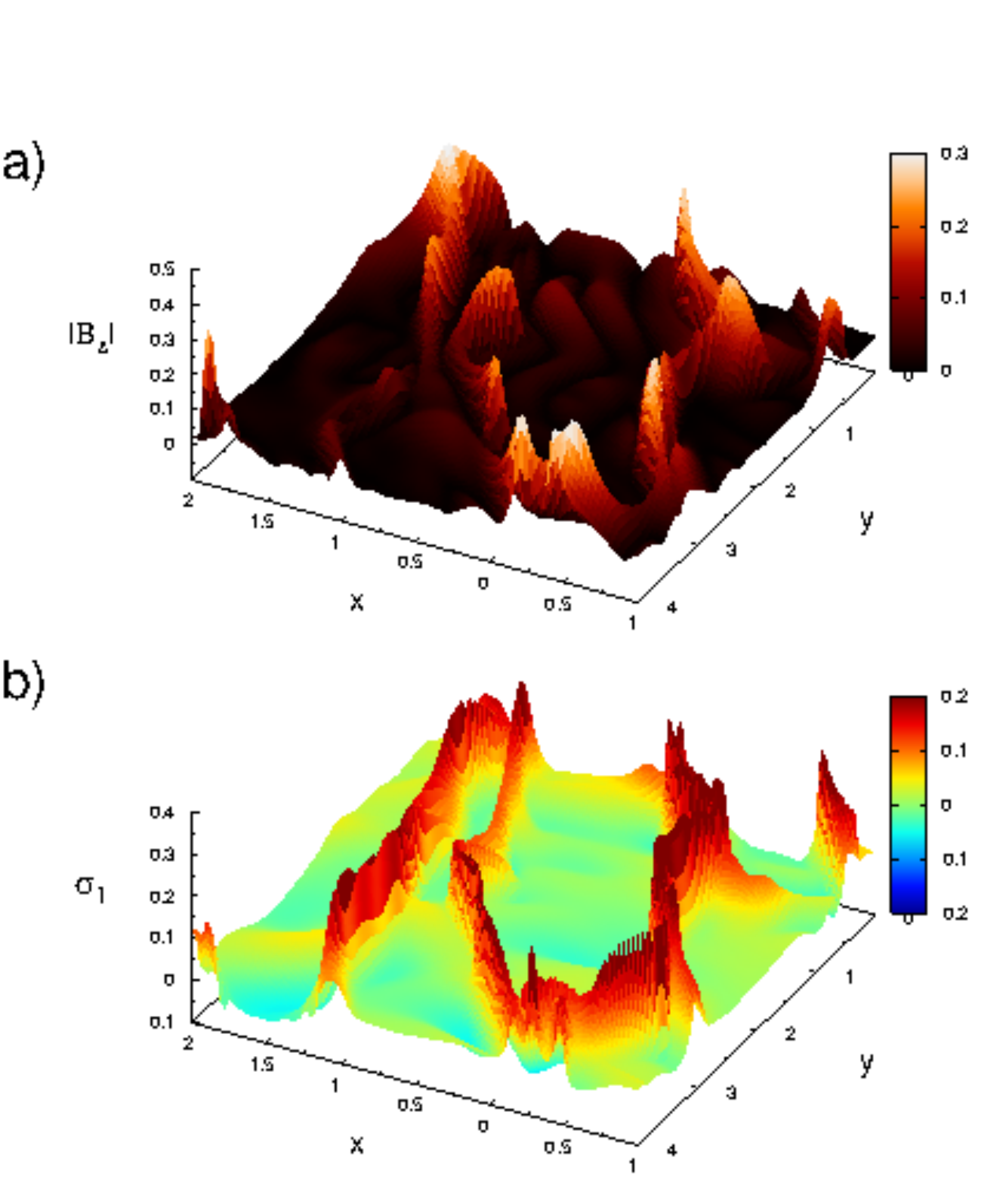}
\end{center}
 \caption{{\bf 3D plots of numerical simulation of compressible convection corresponding to the rectangle regions indicated in Figs. \ref{fig blcs2d}(a)-(b).} {\bf (a)} Modulus of the vertical component of the magnetic field $B_z$ at 
time $t_0$; {\bf (b)} the b-FTLE for $t_0$ and $\tau = -10$ time units. }
\label{fig blcs_sim3d}
 \end{figure}

Figure \ref{fig blcs_sim2d}(a) shows a 2D image of the vertical component of
the magnetic field $B_z$ near the top of the convective layer at a time $t_0$ when the magnetic
energy growth has already saturated. 
This figure shows that the simulated large-scale turbulent dynamo reproduces the pattern of 
convective cells, similar to granulations in the photosphere. In particular, it  shows that the 
network of high concentration of magnetic flux is located in the boundaries between the convective cells, 
corresponding to the intergranular lanes. The elongated convective cells in Figure \ref{fig blcs_sim2d}
 result from the shear. Since there is an imposed large-scale flow in the $y$ direction, it is expected 
that the convective cells would be elongated along this direction. This is more pronounced after a 
strong large-scale magnetic field develops due to the dynamo mechanim.
Figure \ref{fig blcs_sim2d}(b)
shows the b-FTLE computed for the horizontal velocity field $(v_x, v_y)$ at $t_0$ using $\tau=-10$ time units. 
The ridges of high values of b-FTLE in Figure  \ref{fig blcs_sim2d}(b) indicate
the locations of converging plasma flows.
In Figure  \ref{fig blcs_sim2d}(c) we plot the superposition of the thresholded b-FTLE
(with $\sigma_1>0.1$) on top of $B_z$. For the sake of clarity, a 3D view of the rectangle 
regions of Figure  \ref{fig blcs_sim2d} is given in Figure  \ref{fig blcs_sim3d}, where the modulus of $B_z$ is
plotted in comparison with the b-FTLE. Evidently,
Figures  \ref{fig blcs_sim2d} and  \ref{fig blcs_sim3d} render support for the results 
of Figures  \ref{fig blcs2d} and  \ref{fig blcs3d} that there is correspondence of the
network of high magnetic flux concentration to the attracting LCS in the photospheric velocity.

In conclusion, we extended the observational study by \citet{yeates12} of Lagrangian coherent
structures in the photospheric velocity to show that in addition to the correspondence
of the network of quasi-separatrix layers in the repelling LCS demonstrated by \cite{yeates12},
there is a correspondence of the network of high magnetic flux concentration
to the attracting LCS in the photospheric flows, confirmed by both 
observations and numerical simulations. Hence, the repelling and attracting LCS
of the photospheric velocity provide complementary informations of the local dynamics and 
topology of solar magnetic fields. Although the corks coincide with the network of magnetic flux,
they are not the best way to find the transport barriers and the directions of the 
flow convergence. We showed that the attracting LCS are more suitable 
for these tasks because they trace continuous curves (material lines) that are the explicit structures potentially responsible for observed physical patterns and at the same 
time measure the local rate of contraction/expansion in the flow. While 
Eulerian coherent structures provide useful insights of the snapshot of the 
photospheric flows, they fail to reveal the fine structures of the transport
barriers and the local dynamics of the flows that are readily provided by the repelling
and attracting Lagrangian coherent structures.



\acknowledgments
A.C.L.C. acknowledges the support of CNPq, the award of a Marie Curie Fellowship by the European Commission, and the hospitality of Paris Observatory.
ELR acknowledges the financial support of CNPq (Brazil) and FAPESP (Brazil). We thank Petri K\"apyl\"a for providing the
configuration files for the PENCIL CODE MHD simulations.

\clearpage

\end{document}